\documentclass[aps,twocolumn,groupedaddress,superscriptaddress,amsmath,amssymb,prl]{revtex4-2}

\usepackage{amsfonts}
\usepackage{amsmath}
\usepackage{float}
\usepackage{physics}
\usepackage{mathtools}
\usepackage{bbold}
\usepackage{amsthm}

\usepackage[margin=25mm]{geometry}
\usepackage[table,xcdraw]{xcolor}
\usepackage{amsmath}
\usepackage{braket}
\usepackage{hyperref}

\usepackage{tikz}
\usepackage{pgfplots}
\usepackage{tikzscale}
\usepackage{ mathrsfs }
\newcommand{\bqa}{\begin{eqnarray}}
\newcommand{\eqa}{\end{eqnarray}}

\usepackage{comment}
\usepackage{placeins}

\newcommand{\be}{\begin{equation}}
\newcommand{\ee}{\end{equation} }

\newcommand{\flo}[1]{{\color[rgb]{0,0,0.6}{#1}}}

\begin{document}

\title{Iterative Quantum Optimization with Adaptive Problem Hamiltonian}

\author{Yifeng Rocky Zhu}
\affiliation{Physics Department, Blackett Laboratory, Imperial College London, Prince Consort Road, SW7 2AZ, United Kingdom}
\author{David Joseph}
\affiliation{SandboxAQ, Palo Alto}
\author{Cong Ling}
\affiliation{Electrical and Electronic Engineering Department, Imperial College London}
\author{Florian Mintert}
\affiliation{Physics Department, Blackett Laboratory, Imperial College London, Prince Consort Road, SW7 2AZ, United Kingdom}
\affiliation{Helmholtz-Zentrum Dresden-Rossendorf, Bautzner Landstraße 400, 01328 Dresden, Germany}

\begin{abstract}
Quantum optimization algorithms hold the promise of solving classically hard, discrete optimization problems in practice.
The requirement of encoding such problems in a Hamiltonian realized with a finite -- and currently small -- number of qubits, however, poses the risk of finding only the optimum within the restricted space supported by this Hamiltonian.
We describe an iterative algorithm in which a solution obtained with such a restricted problem Hamiltonian is used to define a new problem Hamiltonian that is better suited than the previous one.
In numerical examples of the shortest vector problem, we show that the algorithm with a sequence of improved problem Hamiltonians converges to the desired solution.
\end{abstract} 

\maketitle

\section{Introduction}

The currently available hardware for quantum information processing is getting close to the specifications that are required for the solution of real-world problems~\cite{arute2019quantum}.
As a result of the anticipated ability of quantum computers to break popular cryptographic protocols, a new generation of protocols has been developed~\cite{bernstein2009introduction}.
These post-quantum cryptographic protocols are expected to be secure even if an eavesdropper had access to a fully-functioning quantum computer.
There is a variety of such protocols and the assessment of their security either in terms of security proof or explicit counter-attack is a central goal of the community.

A large class of post-cryptographic protocols is based on the shortest vector problem (SVP) of lattices.
Similar to the prime-factorization problem that RSA~\cite{RSA} is built upon, SVP is also a seemingly simple problem that turns out to be computationally difficult to solve~\cite{BK:LLL}.
It is a discrete optimization problem, defined in terms of a basis of a finite-dimensional vector space. Lattice vectors are obtained by forming linear superpositions of the basis vectors with integer expansion coefficients.
Usually, the security of lattice-based cryptography can be reduced to the problem of finding the shortest, non-zero vector of a lattice -- in the following simply referred to as the {\it shortest vector}.

The computational effort required to find the shortest vector depends on the properties of a basis;
with a {\em good basis} of short vectors that are close-to-orthogonal to each other, the shortest vector can typically be found in practice,
but with a {\em bad basis} of long and close-to-parallel vectors finding the shortest vector is computationally intractable,
even with the largest currently existing classical high-performance computers for lattices with a dimension in the hundreds.
A crytographic protocol can therefore be based on a publicly known bad basis~\cite{BK:LLL}.

Since the shortest-vector problem can be mapped onto a quantum Ising Hamiltonian~\cite{PhysRevA.103.032433},
such that its eigenvectors and eigenvalues correspond to lattice vectors and their squared lengths,
it can readily be formulated as a quantum mechanical algorithm such as an adiabatic algorithm~\citep{aharonov2008adiabatic,farhi2000quantum}, variational quantum eigensolver~\citep{peruzzo2014variational,mcclean2016theory,wecker2015progress,APSW22} or a quantum approximate optimization algorithm (QAOA)~\cite{farhi2014quantum,joseph2021quantum}. 

A crucial issue in all these implementations is that any finite number of qubits allows only for an optimization over a finite range of values of the expansion coefficients.
Even though, there is a minimal number of qubits that guarantees that the shortest vector can be found~\cite{
PhysRevA.103.032433,APSW22}, this number is far out of reach for current and foreseeable technology.
Even in the absence of any imperfections, such as limited gate fidelities or decoherence, realistic sizes of a qubit register would thus result in the risk of finding the shortest vector within a subset of vectors that is not the actual shortest vector.

As we will show here, the qubit requirement to realize a problem Hamiltonian in a sufficiently large Hilbert space based on a bad basis is indeed very stringent.
Finding a reasonably short, but not necessarily the shortest vector within a subset of vectors, however, helps to construct a better basis than the originally used one.
A quantum algorithm with a problem Hamiltonian based on this improved basis then gives access to shorter vectors than the one based on the original problem Hamiltonian.
The resultant iterative improvement of basis and corresponding problem Hamiltonian enables the search for the actual shortest vector even under stringent limitations of available qubits.

\section{Quantum optimization for the shortest vector problem}

A lattice is the collection of points in a $d$-dimensional space given by the linear superpositions $\sum_{i=1}^dn_i {\bf b}_i$ of basis vectors ${\bf b}_i$ with integer expansion coefficients $n_i$.
Any lattice~\flo{\footnote{apart from the case of one-dimensional lattices}} can be represented by infinitely many bases;
given one basis $\{{\bf b}_i\}$, any other basis $\{{\bf a}_i\}$ can be formed in terms of linear combinations
 \be
 {\bf a}_i=\sum_jV_{ij}{\bf b}_j\ ,
 \label{eq:basis1}
 \ee
 where the matrix $V$ has integer elements $V_{ij}$ and determinant $\pm 1$, {\it i.e.} it is unimodular.

A quantum mechanical Hamiltonian $H_P$ representing a lattice can be defined as~\cite{PhysRevResearch.2.013361}
\be
H_P =\sum_{ij=1}^d\left({\bf b}_i{\bf b}_j\right)\ \hat Q_i\hat Q_j
\label{eq:problemHamiltonian}
\ee
where the scalar factors ${\bf b}_i{\bf b}_j$ determine the structure of the lattice, and each of the mutually commuting operators $\hat Q_i$ has an integer spectrum.
Any state that is a mutual eigenvector of all the operators $\hat Q_i$ with corresponding eigenvalues $n_i$ corresponds to a lattice vector $\sum_in_i{\bf b}_i$,
and the associated eigenvalue of $H_P$ is given by the  squared length $\sum_{ij=1}^d {\bf b}_i{\bf b}_jn_in_j$ of this lattice vector.

In practice, the operators $\hat Q_i$ act on a Hilbert space that is the tensor product of $d$ smaller factors, and the indices $i$ indicate the factor that the respective operator acts on non-trivially.
Each of the operators $\hat Q_i$ can be realized in terms of several qubits, and the encoding 
\be
\hat Q_i=\frac{1}{2}\left(\sum_{j=1}^k2^{j-1}\hat Z_{ij}+\mathbb{1}\right)\ ,
\label{eq:qubitexpansion}
\ee
with the Pauli $\hat Z$-operator
achieves a non-degenerate spectrum in the range $[-2^{k-1} +1, 2^{k-1}]$ in terms of $k$ qubits.
With this encoding, the problem Hamiltonian for a $d$-dimensional lattice is an Ising Hamiltonian with $dk$ qubits interacting via a $\hat Z\hat Z$-interaction~\cite{PhysRevA.103.032433}.

Deterministically finding the first excited state of this problem Hamiltonian corresponding to the shortest vector is typically not possible with quantum hardware of the limited, currently available specifications, but a QAOA algorithm
\be
\ket{\Psi}=\exp(-i\beta H_D)\exp(-i\gamma H_P)\ket{\Psi_0}
\label{eq:QAOA}
\ee
of lowest depth~\cite{farhi2020quantum} can realise a state $\ket{\Psi}$ that results in high probabilities to project onto a low-lying eigenstate of $H_P$ upon measurement of the observables $\hat Q_i$ if initial state $\ket{\Psi_0}$, Rabi-angles $\beta$ and $\gamma$ and driver Hamiltonian $H_D$ are chosen suitably~\citep{joseph2021quantum,bittel2021training}.

\section{Limited parameter range}

The problem Hamiltonian in Eq.~\eqref{eq:problemHamiltonian} is defined in terms of a basis of the lattice,
and different choices for this basis will generally result in different Hamiltonians.
Even though any such problem Hamiltonian has the same spectrum -- given by the squared lengths of all the lattice vectors -- problem Hamiltonians defined with different lattice bases have different physical properties, encoded in the scalar factors ${\bf b}_i{\bf b}_j$.

In any practical implementation with the operators $\hat Q_i$ realized in terms of several qubits, such as the construction given in Eq.~\eqref{eq:qubitexpansion}, the resulting problem Hamiltonians are truncated, and their spectra are only subsets of the spectrum of the full problem  Hamiltonian.
This truncation also breaks the equivalence of different problem Hamiltonians and the spectrum of any truncated problem Hamiltonian depends on the underlying lattice basis.

A problem Hamiltonian constructed in terms of a good basis will be such that the shortest vector is associated with eigenvalues of the operators $\hat Q_i$ that have a small magnitude,
because the expansion of the shortest vector in terms of a good basis requires only small expansion coefficients.
Since, however, the expansion of the shortest vector in terms of a bad basis typically requires large expansion coefficients, the operators $\hat Q_i$ need a broad spectrum in order to ensure that the spectrum of the problem Hamiltonian contains the eigenvalue associated with the shortest vector.

\begin{figure}[t]
\centering
\includegraphics[width=0.45\textwidth]{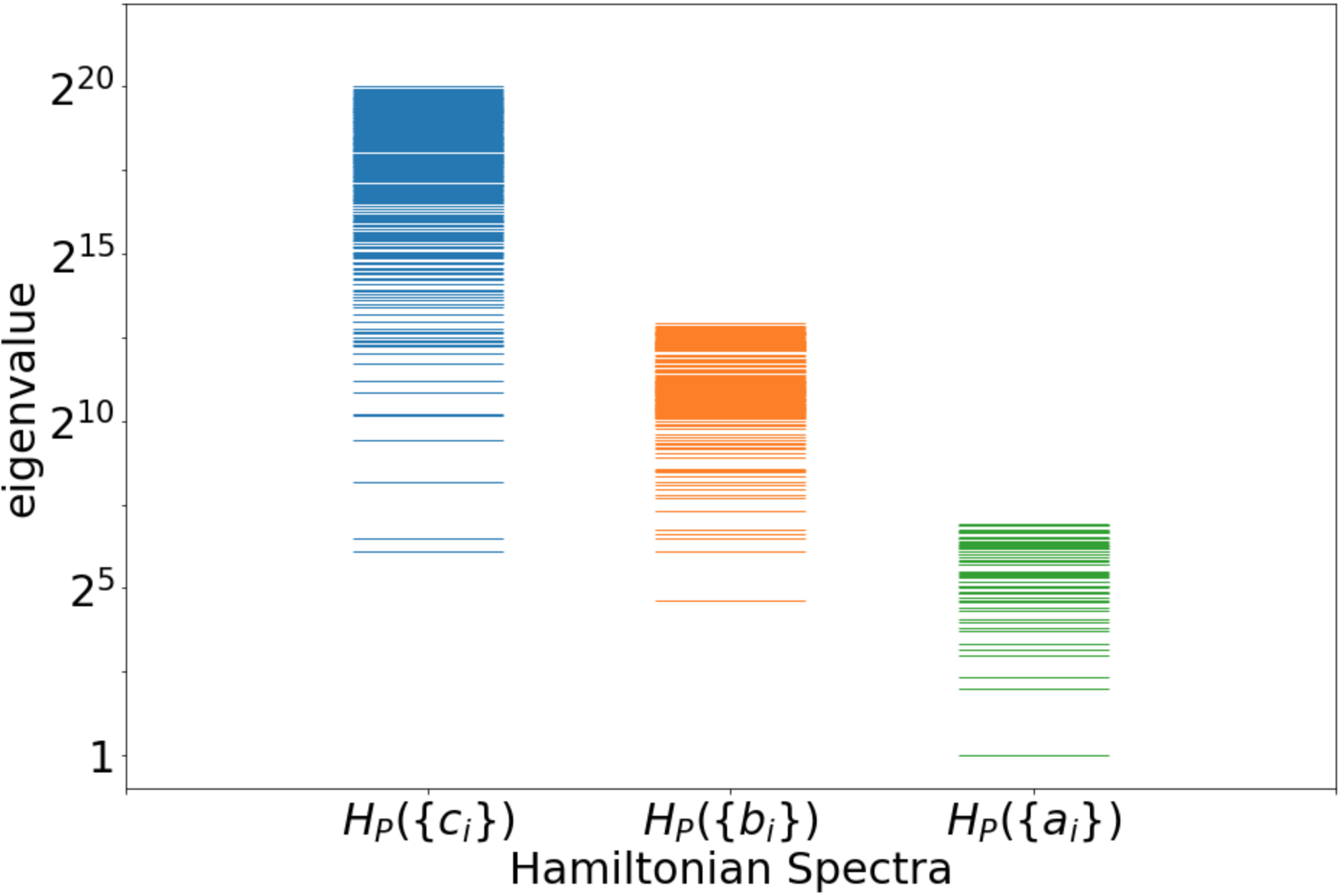}
\caption{\
The blue, orange and green level sets depict the spectra of finite-dimensional problem Hamiltonians (Eq.~\eqref{eq:problemHamiltonian}) for the lattice defined in terms of the bases in Eq.~\eqref{eq:basis_c}, Eq.~\eqref{eq:basis_b} and Eq.~\eqref{eq:basis_a} respectively;
each of the four operators $\hat Q_i$ (Eq.~\eqref{eq:qubitexpansion}) is comprised of $k=2$ qubits.
The green spectrum corresponds to the problem Hamiltonian defined in terms of the shortest basis and it contains the eigenvalue corresponding to the shortest vector of length $1$.
The orange and blue spectra correspond to the problem Hamiltonians defined in terms of the increasingly bad bases, and their lowest eigenvalues lie significantly above the eigenvalue corresponding to the actual shortest vector.}
\label{fig:spectrum}
\end{figure}

This is exemplified for the case of a four-dimensional lattice ${\cal L}$ in Fig.~\ref{fig:spectrum}.
A basis $\{{\bf a}_i\}$ of ${\cal L}$ with vectors of minimal lengths is given by
\begin{subequations}
\begin{alignat}{4}
{\bf a}_1&=[1,0,0,0]\ ,\\
{\bf a}_2&=[0,2,0,0]\ ,\\
{\bf a}_3&=[0,0,3,0]\ ,\\
{\bf a}_4&=[0,0,0,4]\ .
\end{alignat}
\label{eq:basis_a}
\end{subequations}
Two exemplary bases with longer basis vectors are given by
\begin{subequations}
\begin{alignat}{7}
{\bf b}_1&=&[&& 3   &,& 0   &,&  \hspace{.1cm}15  &,& \hspace{.1cm}-12&]&\ ,\\
{\bf b}_2&=&[&& 0   &,& 4   &,&  3   &,& 8&]&\ ,\\
{\bf b}_3&=&[&& \hspace{.1cm}28  &,& \hspace{.1cm}-18 &,&  9   &,& 8&]&\ ,\\
{\bf b}_4&=&[&& 0   &,& 0   &,&  3   &,& -4&]&\ ,
\end{alignat}
\label{eq:basis_b}
\end{subequations}
and
\begin{subequations}
\begin{alignat}{12}
{\bf c}_1&=&[&& \hspace{.1cm}25&,& 78&,&\hspace{.1cm}105&,&\hspace{.1cm}160&]&\ ,\\
{\bf c}_2&=&[&&  \hspace{.1cm}-3&,& 32&,&18&,& 64&]&\ ,\\
{\bf c}_3&=&[&&\hspace{.1cm}53&,&\hspace{.1cm}128&,&\hspace{.1cm}195&,&\hspace{.1cm}264&]&\ ,\\
{\bf c}_4&=&[&&  0&,&  8&,&  9&,& 12&]&\ .
\end{alignat}
\label{eq:basis_c}
\end{subequations}
All three bases represent the same lattice $\mathcal{L}$.
The basis $\{{\bf a}_i\}$ qualifies as a good basis, while $\{{\bf c}_i\}$ is a bad basis, and the basis $\{{\bf b}_i\}$ is clearly better than $\{{\bf c}_i\}$, but substantially worse than $\{{\bf a}_i\}$. 

Fig.~\ref{fig:spectrum} depicts the spectra of the problem Hamiltonians constructed with either of these three bases and each operator $\hat Q_i$ realized in terms of $k=2$ qubits.
Whereas the spectrum of the problem Hamiltonian constructed with the basis $\{{\bf a}_i\}$ (green) covers mostly low-lying states of the spectrum of the full problem Hamiltonian, the spectra of the problem Hamiltonians constructed with the basis $\{{\bf b}_i\}$ (orange) and $\{{\bf c}_i\}$ (blue) cover substantially more high-lying states.
In particular, the lowest non-zero states encoded in these two Hamiltonians are substantially higher than the actual shortest vector.

\begin{figure}[t]
\centering
\includegraphics[width=0.45\textwidth]{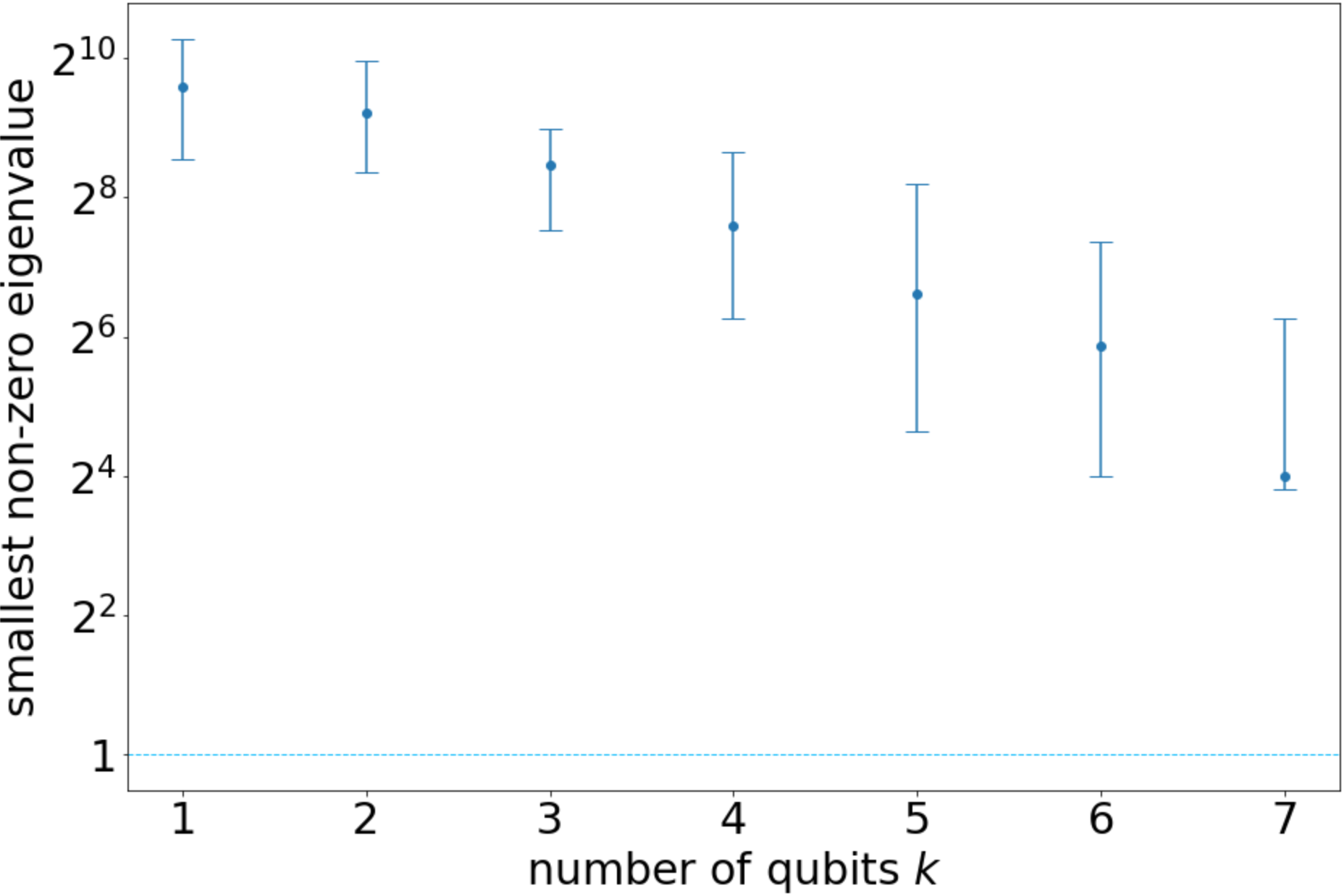}
\caption{
Lowest non-zero eigenvalues of problem Hamiltonians for the four-dimensional lattice defined in Eq.~\eqref{eq:basis_a} as function of the number of qubits $k$ used in the realization of the operators $\hat Q_i$ in Eq.~\eqref{eq:qubitexpansion}.
The dots depict the medians over $100$ problem Hamiltonians constructed in terms of randomly chosen unimodular matrices, and the error bars depict the corresponding $75^{th}$ quantiles.}
\label{fig:scaling}
\end{figure}

A problem Hamiltonian constructed with a bad basis that does include the eigenstate that is associated with the shortest vector would thus require substantially more qubits per lattice dimension.
Fig.~\ref{fig:scaling} depicts the smallest non-zero eigenvalue of problem Hamiltonians realised with different numbers of qubits per lattice dimension as a function of the qubit number $k$.
The dots depict the medians over $100$
bases constructed with randomly chosen unimodular matrices $W=LU$ obtained as product of a lower triangular matrix $L$ and an upper triangular matrix $U$ with unit diagonal elements, and all other non-vanishing elements are chosen randomly from a uniform distribution within the range $[-10,10]$;
the error bars depict the $75^{th}$ quantiles.

The smallest non-zero eigenvalues clearly decrease with the number of qubits, but even with seven qubits, the median is still larger than the shortest vector length (indicated by a light-blue line).
Extrapolating from Fig.~\ref{fig:scaling} suggests that about $10$ qubits per lattice-dimension are required to ensure that the eigenvector associated with the shortest lattice vector is contained in the explicit realization of the problem Hamiltonian. 

Since this qubit requirement seems far out of reach with near-future technology~\citep{preskill2018quantum,bharti2101noisy}, we will present in the following an adaptive algorithm that is based on a gradual improvement of bases and corresponding implementation of problem Hamiltonian.

\section{Adaptive problem Hamiltonian}

The Iterative Quantum Optimization with Adaptive Problem Hamiltonian (IQOAP) algorithm is initialised with the problem Hamiltonian constructed with the bad basis that is publicly available in a cryptographic protocol.
An algorithm such as QAOA that can find a low-lying state in the spectrum of this problem Hamiltonian at least probabilistically produces a lattice vector.
If it is possible to replace one of the basis vectors with this newly obtained vector, while maintaining a basis ({\it i.e.} a set of vectors that spans the complete lattice, or, equivalently a set of vectors related to the original basis via  Eq.~\eqref{eq:basis1} by a unimodular matrix $V$), the basis is updated provided that the new basis vector is shorter than the basis vector that is being dropped.
Independently of whether the basis had been updated or not, the algorithm continues with the above quantum optimization using the problem Hamiltonian constructed with the current basis.

The central advantage of this strategy is that there is no minimal qubit number required to run the quantum optimization.
Even a realization with too few qubits to encode the actual shortest vector can help to find a better basis, which in turn will help to find short vectors at the available qubit count.
One may certainly expect that the rate of convergence depends on the number of utilized qubits, but as we will show in the following, even an implementation with few qubits does typically result in reliable convergence to the actual shortest vector.

In the following discussion, the quantum mechanical part of IQOAP is performed in terms of QAOA with a single step in terms of problem Hamiltonian and driver Hamiltonian each, as given in Eq.~\eqref{eq:QAOA}.
The driver Hamiltonian $H_D=\sum_j\hat X_j$ is given by the collective Pauli $\hat X$, and the Rabi angle $\beta$ is chosen to coincide with the Rabi angle $\gamma$.
The value of $\gamma$ is chosen to minimize the energy expectation value $\bra{\Psi}H_P\ket{\Psi}$. Since this expectation value can be classically evaluated without explicit construction of the state $\ket{\Psi}$ in Eq.~\eqref{eq:QAOA}~\citep{farhi2016quantum,joseph2021quantum}, this minimization can be performed classically without using quantum mechanical computational resources.

Each run of QAOA yields a lattice vector $\textbf{v}$. This is substituted into the basis if the following criteria are met:
\begin{enumerate}
    \item $\textbf{v}$ is shorter than some basis vector $\textbf{b}$,
    \item Replacing $\textbf{b}$ with $\textbf{v}$ preserves the lattice,
    \item If there are multiple eligible vectors to replace, then the longest eligible basis vector is replaced by $\textbf{v}$.
\end{enumerate}
If the above criteria are not met, then the QAOA circuit is repeated with the same problem Hamiltonian.

\begin{figure}[t]
\centering
\includegraphics[width=0.45\textwidth]{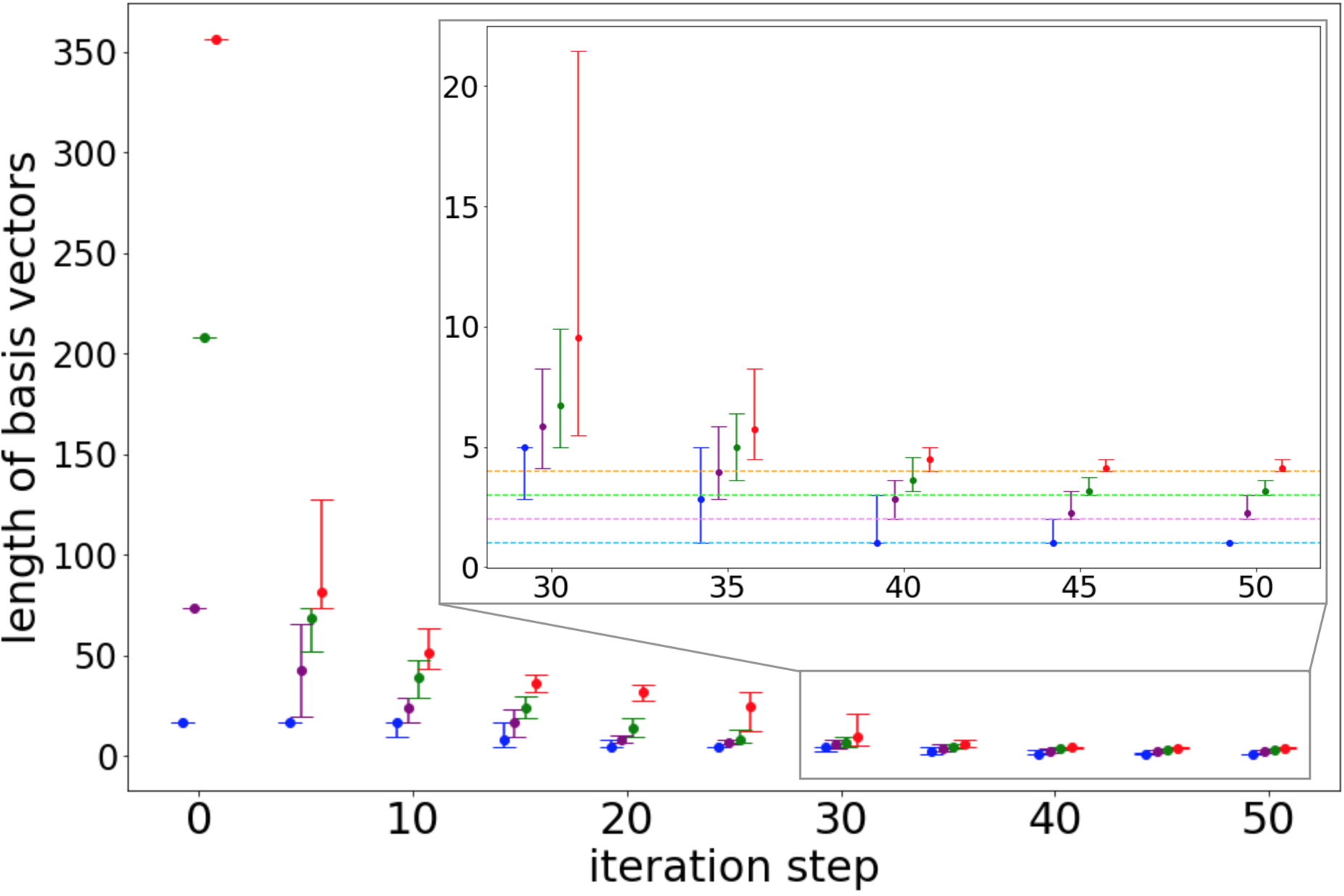}
\caption{
Convergence of the basis during an IQOAP algorithm for the four-dimensional basis defined through Eq.~\eqref{eq:basis_c}.
The $y$-axis shows the lengths of the four basis vectors distinguished by color.
The dots and error bars depict the medians and the $80^{th}$ quantiles over $50$ iterations of the algorithm with fluctuations resulting from the probabilistic nature of QAOA.
The inset depicts a zoom in to the later stage of the algorithm, and the horizontal lines depict the length of the basis vectors given in Eq.~\eqref{eq:basis_a}. 
}
\label{fig:convergence}
\end{figure}
Fig.~\eqref{fig:convergence} depicts an example of how the basis vectors of the four-dimensional lattice defined through Eq.~\eqref{eq:basis_c} decrease as the algorithm progresses.
The problem Hamiltonian is encoded with $k=2$ qubits for each operator $\hat Q_i$ in Eq.~\eqref{eq:qubitexpansion}.

The dots depict the medians over $50$ independent executions of the algorithm and the error bars indicate the $80^{th}$ quantiles.
For reasons of visibility only data for every fifth iteration step is depicted and the four data sets are depicted with different offsets on the $x$-axis in order to avoid overlapping symbols. 
Typically it is possible to update the basis after $2$ to $10$ repetitions of QAOA, and indeed, the lengths of all the four basis vectors decrease rapidly as the algorithm progresses.

The inset shows a zoom in to the later stage of the algorithm with the convergence towards the shortest basis.
After $50$ iterations, the actual shortest vector is found with high probability ($82\%$), and also the other obtained basis vectors coincide with the shortest vectors (Eq.~\eqref{eq:basis_a}) in the vast majority of cases.
 
\begin{figure}[t]
\centering
\includegraphics[width=0.45\textwidth]{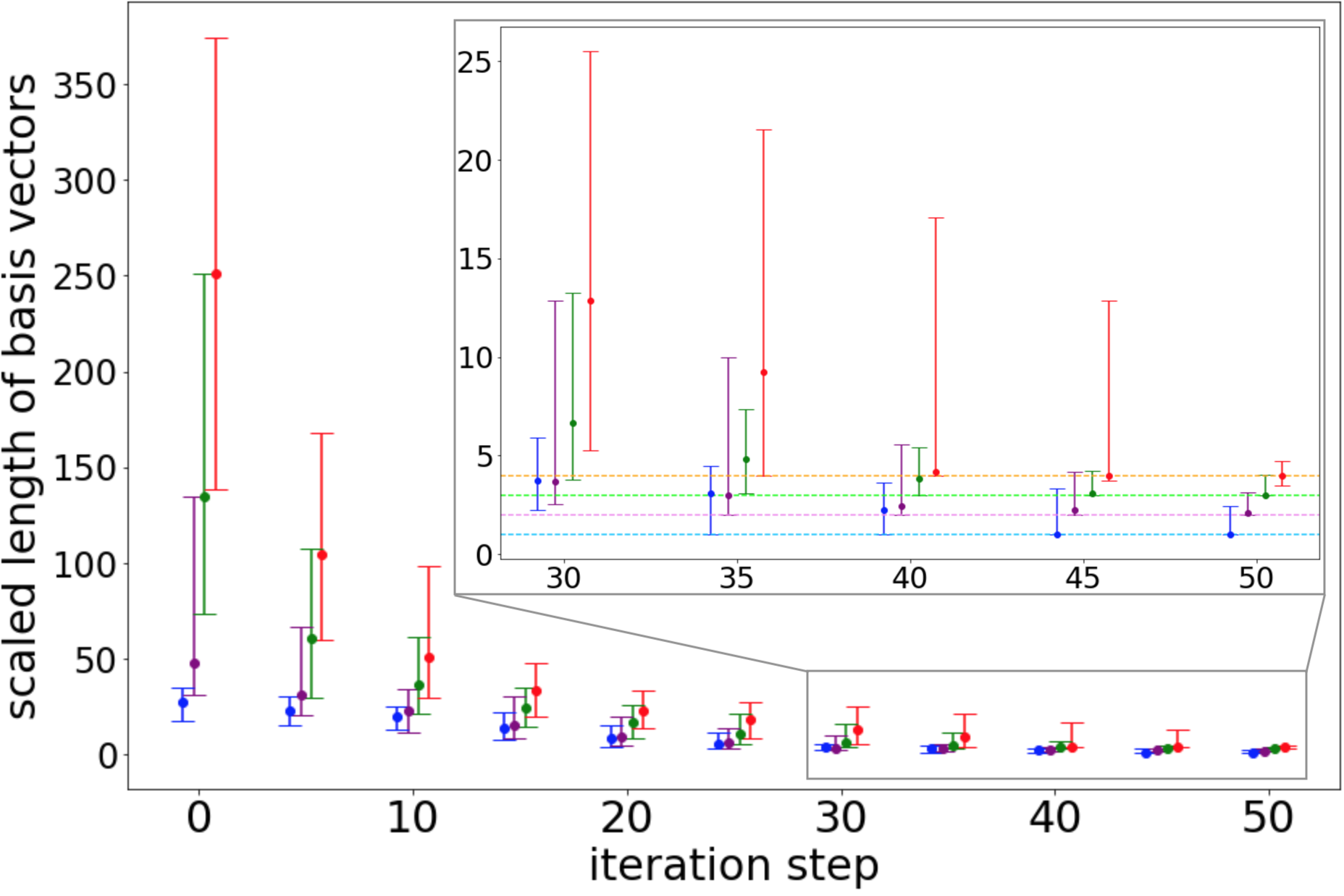}
\caption{Convergence of the IQOAP algorithm with $50$ randomly chosen four-dimensional lattices.
Since the lengths of the shortest vectors are different for different lattices, the $y$-axis depicts the lengths of the basis vectors normalised to the lengths of the shortest basis vectors with a scaling factor of $j=1,2,3,4$.
The medians and $80^{th}$ quantiles are depicted by dots and error bars.}
\label{fig:specbases}
\end{figure}

These convergence properties are indeed not specific to this particular lattice, as shown in
Fig.~\eqref{fig:specbases} which depicts the convergence of the algorithm for different four-dimensional lattices.
The rate of convergence is very similar to that shown in Fig.~\eqref{fig:convergence}; only the fluctuations around the medians (depicted by the error bars) are a bit larger.
Also here, the actual shortest vector is found within $50$ iterations in the vast majority of cases, highlighting that the algorithm converges reliably independent of the properties of the underlying lattice.

\section{Conclusions}

The ability to update the problem Hamiltonian during the progress of the algorithm opens up a new avenue of combining classical and quantum mechanical elements in an algorithm.
Whereas many current hybrid algorithms such as VQE ~\citep{peruzzo2014variational,mcclean2016theory,wecker2015progress,APSW22} have quantum and classical components -- the quantum mechanical evaluation of a function and its classical minimization -- that are independent of each other,
the classical and quantum mechanical aspects in IQOAP are closely intertwined in that extracting classical information from QAOA~\cite{farhi2014quantum,joseph2021quantum} (or a similar algorithm) and classically updating the basis and corresponding problem Hamiltonian creates a modified problem to be solved by quantum mechanical means.
The updated problem in turn gives access to classical information of increased relevance, and this interplay of quantum and classical elements then results in the algorithm's convergence.

The weight of classical and quantum mechanical components in the algorithm can be readily shifted to either side.
An increasing number of available qubits gives access to broader spectra of the truncated problem Hamiltonians, which reduces the number of classical basis-updates before the shortest vector is found.
The effort on the quantum mechanical side of the algorithm can be reduced if classical lattice reduction algorithms~\cite{BK:LLL} are employed together with the basis updates.

While certainly beyond the capabilities of current quantum technologies, one can also envision a more coherent version of this algorithm, in which there is no classical readout during the algorithm, but a problem Hamiltonian conditioned on the current state of the algorithm executed so far is being implemented.

The multiple possibilities to expand the present algorithm in terms of classical or quantum mechanical components make this algorithm sufficiently versatile for applications also beyond the presently discussed lattice problems, such as the closest vector problem (CVP)~\cite{BK:LLL} and learning with errors (LWE)~\cite{Re09}.

\section*{Acknowledgements} 
This work was supported in part by the Engineering and Physical
Sciences Research Council (EPSRC) under Grant No. EP/S021043/1.

\bibliography{IQOAP}

\end{document}